\documentclass[doublecol]{epl2}

\begin{document}

\title{On return-volatility correlation in financial dynamics}
\author{J. Shen and B. Zheng\footnote{corresponding author; email: zheng@zimp.zju.edu.cn}}

\institute{Zhejiang University, Zhejiang Institute of Modern
Physics, Hangzhou 310027, PRC}


\abstract {With the daily and minutely data of the German DAX and
Chinese indices, we investigate how the return-volatility
correlation originates in financial dynamics. Based on a retarded
volatility model, we may eliminate or generate the return-volatility
correlation of the time series, while other characteristics, such as
the probability distribution of returns and long-range
time-correlation of volatilities etc., remain essentially unchanged.
This suggests that the leverage effect or anti-leverage effect in
financial markets arises from a kind of feedback return-volatility
interactions, rather than the long-range time-correlation of
volatilities and asymmetric probability distribution of returns.
Further, we show that large volatilities dominate the
return-volatility correlation in financial dynamics.}

\maketitle

Financial markets are complex systems with many-body interactions.
The possibility of accessing large amounts of historical financial
data have spurred the interest of physicists, to analyze the
financial dynamics with physical concepts and methods. Some
"stylized facts" of the financial markets are revealed
\cite{man95,gop99,liu99,bou01,gab03,ren03,qiu06,she09}. Different
models and theoretical approaches have been proposed to describe and
reproduce the features of the financial dynamics
\cite{cha97,lux99,sta99,con00,egu00,muz00,gia01,cha01,lou01,kra02,zhe04b,zhe04,zho05,ren06a}.

A complex system is often characterized by time correlations and
spatial correlations. A famous stylized fact of the financial
dynamics is the "volatility clustering", i.e., the long-range
time-correlation of volatilities, though the price return itself is
short-range correlated in time \cite{gop99,liu99,gia01}. Meanwhile
recent researches are concerned with the cross-correlations between
different stocks and their statistical properties in different stock
markets \cite{lal99,ple99,mor00,gop01,ple02a,uts04,pan07a,she09}. To
further understand the financial dynamics, one may consider the
higher-order time-correlations. It was first observed by Black
\cite{bla76,cox76} that past negative returns increase future
volatilities, i.e., the return-volatility correlation is {\it
negative}. This is the {\it leverage effect} in financial markets.
In the past years many literatures have been devoted to the leverage
effect, and various relevant correlation coefficients have been
measured within GARCH-like models \cite{hau91,cam92,cam97,bek00}.
Recently Bouchaud et al quantitatively computed the
return-volatility correlation function with the daily data of
several financial markets, and observed that it decays by an
exponential law \cite{bou00,bou01}. More recently, Zheng et al
discovered a {\it positive} return-volatility correlation in Chinese
financial markets \cite{qiu06,qiu07}, i.e., the so-called {\it
anti-leverage effect}. Further, it is shown that both the leverage
effect in German markets and the anti-leverage effect in Chinese
markets can be detected on both {\it daily} and {\it minutely} time
scales \cite{qiu06,qiu07}.

How does the return-volatility correlation originate in financial
dynamics? The economic interpretation of this phenomenon is still
controversial \cite{hau91,bek00}. According to Black, a price drop
increases the risk of a company to go bankrupt, and its stock
therefore becomes more volatile. This induces the leverage effect.
Different models have been proposed to explain the leverage effect
with certain success
\cite{bai96,bou01,zum04,gir04,mas06,tan06,rui08}. The retarded
volatility model is a good example \cite{bou01}. The core thought of
this model is that the reference price used to set the scale for
price updates is not the instantaneous price but rather a moving
average of the price over a past period of time. In fact, the
retarded volatility model may generate both the leverage and
anti-leverage effects by selecting appropriate coupling parameters
$K(t)$ \cite{bou01,qiu06,qiu07}. More recently, there have been
discussions whether the long-range time-correlation of volatilities
may play an important role in the origination of the leverage effect
\cite{gir04,ahl07,rom08}. Especially, it is argued that {\it both}
the long-range time-correlation of volatilities and asymmetric
probability distribution of returns are {\it necessary} in order to
have a leverage effect \cite{rom08}.

By the definition of the return-volatility correlation function, the
long-range time-correlation of volatilities and asymmetric
probability distribution of returns {\it together} may indeed induce
or alter the leverage effect or anti-leverage effect. The question
is only whether it is really a dominating mechanism {\it in
financial markets}. In fact, the shuffling procedure (i.e., randomly
changing the time ordering of returns) naturally destroy not only
the long-range time-correlation of volatilities, but also the
return-volatility correlation. Similarly, randomly changing the sign
of the return removes also both the asymmetry of the probability
distribution of returns and the return-volatility correlation. These
arguments provided in Ref. \cite{rom08} should not be the evidences
in this respect.

In this paper, we investigate how the leverage and anti-leverage
effects originate in financial markets, with the daily and minutely
data of the German DAX and Chinese indices, and based on a retarded
volatility model. The German DAX exhibits a standard leverage
effect. To the best of our knowledge, the Chinese financial market
is the unique one with an anti-leverage effect. We collect the daily
data of the German DAX from 1959 to 2009 with 12\ 407 data points,
and the minutely data from 1993 to 1997 with 360\ 000 data points.
The daily data of the Shanghai Index are from 1990 to 2009 with 4\
482 data points, and the minutely data are from 1998 to 2006 with
95\ 856 data points. The daily data of the Shenzhen Index are from
1991 to 2009 with 4\ 435 data points, and the minutely data are from
1998 to 2003 with 50\ 064 data points. The minutely data are
recorded every minute in the German DAX, while every 5 minutes in
the Chinese indices. A working day is about 450 minutes in Germany
while exactly 240 minutes in China. The dynamic behavior of the
Shenzhen Index is similar to that of the Shanghai Index. Sometimes
we show only the results of the latter.

We denote the price of a stock index at time $t'$ as $P(t')$, then
its logarithmic price return over a time interval $\Delta{t}$ is
\begin{equation}
R(t',\Delta{t})\equiv{\ln{P(t'+\Delta t)}-\ln{P(t')}}. \label{e1}
\end{equation}
For comparison of different financial indices, we introduce the
normalized return
\begin{equation}
r(t')\equiv [{R(t',\Delta{t})-<R(t',\Delta{t})>]/\sigma}, \label{e2}
\end{equation}
where $<\ldots>$ represents the average over time $t'$, and $\sigma
= {\sqrt{<R^{2}>-<R>^{2}}}$ is the standard deviation of
$R(t',\Delta{t})$. Following Ref. \cite{bou01}, the
return-volatility correlation function is defined by
\begin{equation}
L(t)=<r(t')|r(t'+t)|^{2}>/Z,
 \label{e3}
\end{equation}
with $Z=<|r(t')|^{2}>^{2}$. In general, $L(t)$ also depends on
$\Delta{t}$. In this paper, we take $\Delta{t}$ to be the minimum
time unit, i.e., one day for the daily data, and one minute and five
minutes for the minutely data of the German DAX and Chinese indices
respectively.

\begin{figure}
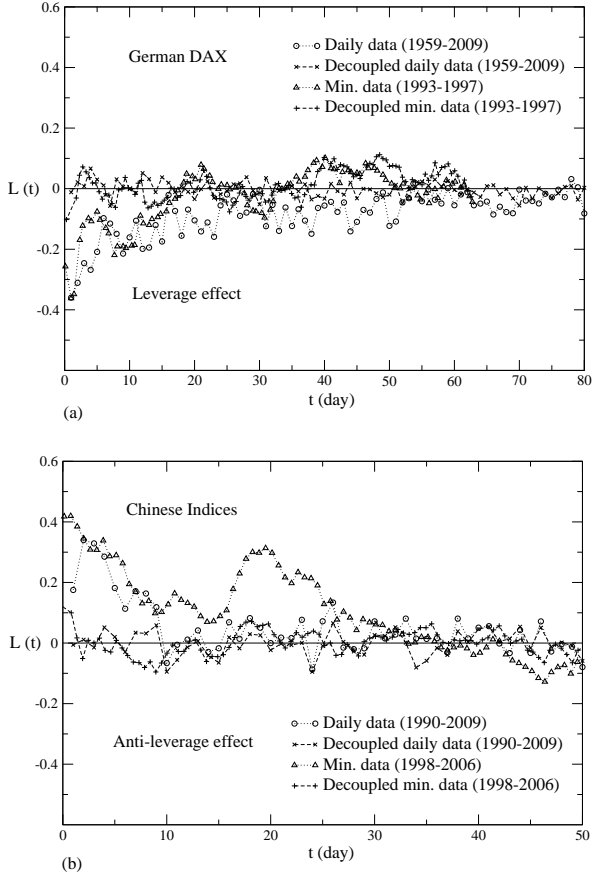

\onefigure[scale=0.32]{Leverage_DAX.eps} \vspace{0.4cm}
\onefigure[scale=0.32]{Leverage_CHN.eps} \caption{The
return-volatility correlation is displayed for the daily and
minutely data of both the German DAX and Chinese indices. After the
decoupling interactions are introduced, the leverage effect of the
German DAX in (a) and the anti-leverage effect of the Chinese
Indices in (b) are eliminated.} \label{f1}
\end{figure}

In fig. \ref{f1}, the return-volatility correlation function is
displayed for the daily and minutely data of both the German DAX and
Chinese indices. $L(t)$ of the German DAX shows negative values up
to at least 20 days. This is the well-known leverage effect. $L(t)$
of the Chinese indices (the average of the Shanghai Index and
Shenzhen Index) takes positive values at least up to 10 days.  That
is the so-called anti-leverage effect \cite{qiu06,qiu07}. For the
minutely data, the original return-volatility correlation functions
are rather noisy due to the high-frequency mode. To extract the
dynamic behavior of the slow mode, we have performed an average over
a 4-day window. This is the finding in Refs. \cite{qiu06,qiu07}.

\begin{figure}
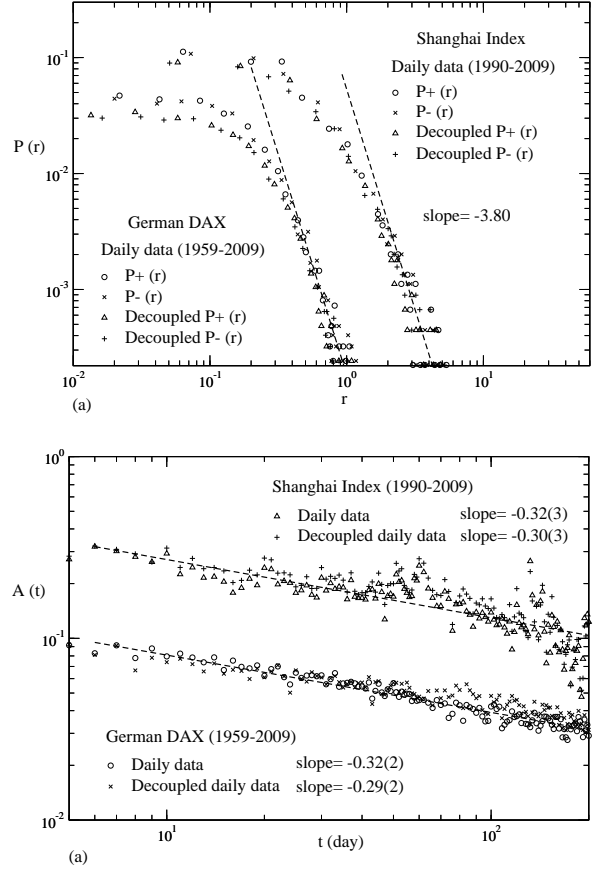

\onefigure[scale=0.32]{PDF_DAY.eps} \vspace{0.4cm}
\onefigure[scale=0.32]{AutoCorrelation_dayvolatility.eps}
\caption{(a) Probability distributions $P_{\pm}(r)$ of positive and
negative returns are displayed for the daily data of the German DAX
and Shanghai Index. $P_{\pm}(r)$ remain almost unchanged after the
decoupling interactions are introduced. For clarity, $P_{\pm}(r)$ of
the German DAX have been shifted to the left by 4.5 units. (b) The
time-correlation function of volatilities is displayed for the daily
data of both the German DAX and Shanghai Index. The long-range
time-correlation remains after the decoupling interactions are
introduced.} \label{f2}
\end{figure}

In fig. \ref{f2} (a), the probability distributions of positive and
negative returns are plotted for the daily data of both the German
DAX and Chinese indices on a log-log scale. Obviously, the positive
and negative tails are not asymmetric. The exponent governing the
power-law decay is about 3.80 \cite{qiu07}. Similar behavior is also
observed for the minutely data. In other words, neither the leverage
effect of the German DAX nor the anti-leverage effect of the Chinese
indices is generated by the asymmetric probability distribution of
returns together with the long-range time-correlations of
volatilities as suggested in Ref. \cite{rom08}.

We now consider a retarded volatility model, which assumes the price
change
\begin{equation}
dP(t')=[P(t')-\sum^{\infty}_{t=1}K(t)dP(t'-t)]\sigma(t')\epsilon(t'),
\label{e4}
\end{equation}
where $\epsilon(t')$ is a Gaussian white noise and $\sigma(t')$ is
the reference volatility \cite{bou01,qiu06}. Usually, $P(t')$ is
with a background, then $dP(t'-t)/P(t')\approx dP(t'-t)/P(t'-t)$.
Keeping in mind that $r(t')\approx dP(t')/P(t')$, Eq. (\ref {e4})
leads to
\begin{equation}
r(t')=[1-\sum^{\infty}_{t=1}K(t)r(t'-t)]\sigma(t')\epsilon(t').
\label{e5}
\end{equation}
Following Ref. \cite{bou01}, one may approximately derive that the
return-volatility correlation function $L(t)$ is about $-2K(t)$ if
$\sigma(t')$ is the order of 1. More generally,
\begin{equation}
K(t)=-\frac{C}{2}L(t), \label{e6}
\end{equation}
with $C$ being a positive constant. Simply speaking, the feedback
return-volatility interaction described by $K(t)$ in Eq. (\ref{e5})
could generate a non-zero $L(t)$ according to Eq. (\ref{e6}). Is
this the real dynamic mechanism in financial markets?

Our thought is to introduce a decoupling return-volatility
interaction to eliminate the non-zero return-volatility correlation
from the time series of returns in real financial markets. If other
characteristics of the time series remain unchanged, one may catch
to some extent how the leverage and anti-leverage effects originate.
For this purpose, we reformulate Eq. (\ref{e5}) as
\begin{equation}
r_{0}(t')=[1+\sum^{\infty}_{t=1}K(t)r(t'-t)]r(t'), \label{e7}
\end{equation}
where $r(t')$ is the original return of real financial markets, and
$r_{0}(t')$ is the decoupled one. Here the decoupling interaction
described by $K(t)$ in Eq.~(\ref{e7}) is just with an opposite sign
compared with the feedback interaction in Eq.~(\ref{e5}). In
practical simulations, we first calculate the return-volatility
correlation function $L(t)$ from $r(t')$, then determine the
parameter $K(t)$ according to Eq.~(\ref{e6}). Our finding is that
the leverage or anti-leverage effect can be eliminated by adjusting
the constant $C$ in Eq. (\ref{e6}) appropriately. In fig. \ref{f1},
$L(t)$ calculated from the decoupled return $r_{0}(t')$ is shown.
Obviously, it fluctuates around zero in all cases, i.e., for the
daily and minutely data of both the German DAX and Chinese indices.
To obtain the decoupled data in fig. \ref{f1}, the constant $C$ is
valued around 0.1 to 0.4 accordingly.

Now it is important to have a survey whether the decoupling
interaction in Eq. (\ref{e7}) changes other characteristics of the
time series $r(t')$. The observation is that the decoupling
interaction is only a {\it perturbation}, in the sense that
$K(t)|r(t'-t)|\ll 1$ and $\sum^{\infty}_{t=1}K(t)r(t'-t)\ll 1$.
Practically, we count the number of data with $K(t)|r(t'-t)|> 1$,
and it is within 5 percent. Particularly, the interacting factor
$1+\sum^{\infty}_{t=1}K(t)r(t'-t)$ almost never changes the sign of
the return, and it only alters the volatility. As a result, except
for the return-volatility correlation, other characteristics of
$r(t')$ remain unchanged.

In fig. \ref{f2} (a), the probability distributions $P_\pm(r)$ of
positive and negative returns are displayed for both the original
daily returns $r(t')$ and decoupled daily returns $r_0(t')$ of the
German DAX and Shanghai Index. Clearly, the decoupling interaction
in Eq. (\ref{e7}) does not modify either the power-law tails or the
shapes of $P_\pm(r)$. Similar results are also observed for the
minutely data. Additionally, we have calculated the mean value
$<r_0(t')>$ for the decoupled daily and minutely returns, and it is
the order of $10^{-3}$ to $10^{-5}$, negligibly small.

The time-correlation function of volatilities is defined as
\begin{equation}
A(t)=[<|r(t')||r(t'+t)|>-<|r(t')|>^{2}]/A_0, \label{e11}
\end{equation}
and $A_0=<|r(t')|^{2}>-<|r(t')|>^{2}$. It is well known that the
volatility in financial dynamics is long-range correlated in time,
i.e., $A(t)$ decays by a power law. In fig. \ref{f2} (b), $A(t)$ is
plotted for both the original daily volatilities and decoupled daily
volatilities of the German DAX and Shanghai Index. A power-law
behavior $A(t)\sim t^{-\beta}$ is detected, both before and after
the decoupling interaction in Eq. (\ref{e7}) is introduced. The
exponent $\beta$ remains almost the same after the leverage or
anti-leverage effect is removed. For example, one estimates
$\beta=0.32(2)$ and $0.32(3)$ for the original daily data of the
German DAX and Shanghai Index, and $\beta=0.29(2)$ and $0.30(3)$ for
the decoupled data \cite{qiu07}.

Because of the intra-day pattern, $A(t)$ calculated with the
minutely data shows a periodic oscillation, and this behavior also
exists for the decoupled minutely data. We remove this intra-day
pattern, e.g., with the procedure in Ref. \cite{liu99,qiu07}, and
estimate the exponents $\beta=0.39(3)$ and $0.33(2)$ for the
original minutely data of the German DAX and Shanghai Index, and
$\beta=0.35(3)$ and $0.34(2)$ for the decoupled data. The values of
$\beta$ of the German DAX are less accurate, because the length of a
working day changes from time to time in Germany.

We could further explore important properties of the financial
dynamics, e.g., the persistence probability of returns or
volatilities. The persistence probability of volatilities $P_{+}(t)$
(or $P_{-}(t)$) is defined as the probability that
$|r(t'+\widetilde{t})|$ has always been above (or below) $|r(t')|$
in time $t$, i.e., $|r(t'+\widetilde{t})|>|r(t')|$ (or
$|r(t'+\widetilde{t})|<|r(t')|$) for all $\widetilde{t}<t$. The
average is taken over $t'$. In general, $P_{-}(t)$ obeys a universal
power-law behavior, $P_{-}(t)\sim t^{-\theta_{p}}$
\cite{ren03,ren05}. In fig. \ref{f4} (a), the persistence
probability $P_{-}(t)$ of volatilities is plotted for both the
original and decoupled daily data of the German DAX and Shanghai
Index. All curves of $P_{-}(t)$ exhibit a nice power-law behavior.
After eliminating the leverage or anti-leverage effect, the exponent
$\theta_{p}$ remains the same. The fact $0.5<\theta_p<1$ indicates
that the volatility is long-range correlated in time.

From our survey above, we conclude that either the leverage effect
or anti-leverage effect does not originate from the asymmetric
probability distribution of returns and long-range time-correlation
of volatilities. The return-volatility correlation is a rather
independent property of the financial dynamics. In fact, with the
retarded volatility model in Eq. (\ref{e5}), one can generate the
leverage effect or anti-leverage effect in a time series,
independent of the probability distribution of returns and
time-correlation of volatilities.

For example, we may take $K(t)=m\ exp(-t/\tau)$
\cite{bou01,qiu06,qiu07}. A positive $K(t)$ induces a leverage
effect, while a negative $K(t)$ leads to an anti-leverage effect
\cite{bou01,qiu06}. $\sigma(t')$ in Eq.(\ref{e5}) is the reference
volatility we need to input, and it can be generated, e.g., by a
Gaussian random process, the EZ model or a dynamic interacting
herding model \cite{egu00,zhe04b,zhe04}. All three models produce
symmetric probability distributions of returns with different
functional forms. The EZ model yields a power-law tail in the
probability distribution of returns, but without the long-range
time-correlation of volatilities. The dynamic interacting herding
model exhibits both a power-law tail in the probability distribution
of returns and the long-range time-correlation of volatilities. In
fig. \ref{f4} (b), the leverage effects from the simulations with
Eq.(\ref{e5}) are presented, in comparison with that of the German
DAX. Here we have chosen $m=0.1$ and $\tau=40$. The anti-leverage
effect can be produce similarly with a negative $m$.

Is the leverage or anti-leverage effect uniformly distributed in the
time series of the financial dynamics, or essentially dominated by
large volatilities? This is important for deeper understanding of
the leverage and anti-leverage effects. In fig. \ref{f6}, the
return-volatility correlation calculated under the conditions
$|r(t')|<2\sigma$, $|r(t')|<8\sigma$, and with all $r(t')$, are
displayed for the daily data of the German DAX and Chinese indices.
Here $\sigma=1$ is the standard deviation of the normalized return
$r(t')$. For the German DAX, the leverage effect becomes very weak
for $|r(t')|<2\sigma$. In other words, the leverage effect is mainly
contributed by the volatilities $|r(t')|>2\sigma$. Since there are
only 7 volatilities larger than $8\sigma$, the curve of $|r(t')|<
8\sigma$ is not too different from that of all $r(t')$. For the
Chinese indices, the anti-leverage effect is essentially dominated
by the volatilities $|r(t')|>8\sigma$, because the curve of $|r|<
8\sigma$ already fluctuates around 0. Briefly speaking, either the
leverage effect of the German DAX or the anti-leverage effect of the
Chinese indices is dominated by the large volatilities, at least for
the daily data. Since the Chinese financial market is more volatile,
such a phenomenon looks more prominent. For the minutely data, it is
somewhat complicated, and small volatilities may also contribute in
some cases. Further, we could investigate the different
contributions of the large volatilities induced by external events
and generated intrinsically by the dynamics itself. Details of this
kind will be presented elsewhere.

In summary, with the retarded volatility model, we can eliminate the
leverage effect of the German DAX and anti-leverage effect of the
Chinese indices on both daily and minutely time scales, while other
characteristics of the time series, such as the probability
distribution of returns, time-correlation and persistence
probability of returns and volatilities etc, remain essentially
unchanged. In addition, the probability distribution of returns of
the German DAX and Chinese indices are not asymmetric before and
after eliminating the leverage or anti-leverage effect. These
results suggest that at least for the German DAX and Chinese
indices, the leverage or anti-leverage effect in financial markets
arises from a kind of feedback return-volatility interactions,
rather than the long-range time-correlation of volatilities and
asymmetric probability distribution of returns. Finally, we show
that the leverage effect of the German DAX and anti-leverage effect
of the Chinese indices are dominated by large volatilities. For the
data set analyzed in Ref. \cite{rom08} (not reachable for us), one
may follow the approach in this paper to clarify how the leverage
effect originates.

{\bf Acknowledgements:} This work was supported in part by NNSF
(China) under grant No. 10875102 and 10325520.

%\bibliographystyle{epl}
%\bibliography{eco2,zheng}

\begin{figure}
\onefigure[scale=0.32]{PersistenceProbability_volatility.eps}
\vspace{0.4cm} \onefigure[scale=0.32]{ModelLeverage.eps}
\caption{(a) The persistence probability $P_{-}(t)$ of volatilities
is plotted for both the original and decoupled daily data of the
German DAX and Shanghai Index. (b) The leverage effects generated by
the retarded volatility model in Eq. (\ref{e5}) with different kinds
of reference volatilities, i.e., the Gaussian random process, EZ
model and dynamic interacting herding model. All three simulations
reproduce the leverage effects similar to that of the German DAX.}
\label{f4}
\end{figure}

\begin{figure}
\onefigure[scale=0.32]{RDAXDAY5909_thresholdleverage.eps}
\vspace{0.4cm}
\onefigure[scale=0.32]{RCHNDAY9009_thresholdleverage.eps}
\caption{The return-volatility correlation of the daily data of the
German DAX (a) and Chinese indices (b), calculated under the
conditions $|r(t')|<2\sigma$, $|r(t')|<8\sigma$, and with all
$r(t')$. $\sigma=1$ is the standard deviation of the normalized
return $r(t')$.} \label{f6}
\end{figure}

\end{document}